\documentclass[11pt,fleqn]{article}
\usepackage{amsfonts,amssymb,latexsym,cite}
\usepackage{epsf,graphicx}

\newcommand{\bls}[1]{\renewcommand{\baselinestretch}{#1}}

\def\noi{\noindent}

\newcommand{\Title}[1]{\noi {{\Large\bf #1}}\\[1ex]}

\def\Aunames#1{\noi{\bf #1}}
\def\auth#1{${}^{#1}$}
\def\Addresses#1{\medskip\noi \protect
    \begin{description}\itemsep -3pt {\it #1} \end{description}}
\def\addr#1#2{\item[${}^{#1}$]{\it #2}}

\newcommand{\Abstract}[1]{\vskip 2mm \begin{center}
        \parbox{16.4cm}{\small\noi #1} \end{center}\medskip}

\def\email#1#2{\footnotetext[#1]{e-mail: #2}\addtocounter{footnote}{1}}

\def\nq{\hspace*{-1em}}
\def\nqq{\hspace*{-2em}}
\def\nhq{\hspace*{-0.5em}}

\def\cm{\hspace*{1cm}}
\def\inch{\hspace*{1in}}

\def\Acknow#1{\subsection*{Acknowledgment} #1}

\def\Jl#1#2{#1 {\bf #2},\ }

\def\ApJ#1 {\Jl{Astroph. J.}{#1}}
\def\CQG#1 {\Jl{Class. Quantum Grav.}{#1}}
\def\DAN#1 {\Jl{Dokl. AN SSSR}{#1}}
\def\GC#1 {\Jl{Grav. Cosmol.}{#1}}
\def\GRG#1 {\Jl{Gen. Rel. Grav.}{#1}}
\def\JETF#1 {\Jl{Zh. Eksp. Teor. Fiz.}{#1}}
\def\JETP#1 {\Jl{Sov. Phys. JETP}{#1}}
\def\JHEP#1 {\Jl{JHEP}{#1}}
\def\JMP#1 {\Jl{J. Math. Phys.}{#1}}
\def\NPB#1 {\Jl{Nucl. Phys. B}{#1}}
\def\NP#1 {\Jl{Nucl. Phys.}{#1}}
\def\PLA#1 {\Jl{Phys. Lett. A}{#1}}
\def\PLB#1 {\Jl{Phys. Lett. B}{#1}}
\def\PRD#1 {\Jl{Phys. Rev. D}{#1}}
\def\PRL#1 {\Jl{Phys. Rev. Lett.}{#1}}

\def\al{&\nhq}
\def\lal{&&\nqq {}}
\def\eq{Eq.\,}
\def\eqs{Eqs.\,}
\def\beq{\begin{equation}}
\def\eeq{\end{equation}}
\def\bear{\begin{eqnarray}}
\def\bearr{\begin{eqnarray} \lal}
\def\ear{\end{eqnarray}}
\def\earn{\nonumber \end{eqnarray}}
\def\nn{\nonumber\\ {}}

\def\nnn{\nonumber\\ \lal }
\def\nnnv{\nonumber\\[5pt] \lal }

\def\eql{\al =\al}

\def\dst{\displaystyle}

\def\fracd#1#2{{\dst\frac{#1}{#2}}}

\def\Half{{\fracd{1}{2}}}

\def\e{{\,\rm e}}
\def\d{\partial}

\def\sign{\mathop{\rm sign}\nolimits}

\def\const{{\rm const}}
\def\eps{\varepsilon}

\def\then{\ \Rightarrow\ }

\newcommand{\vars}[1]{\left\{\begin{array}{ll}#1\end{array}\right.}

\def\mn{_{\mu\nu}}
\def\MN{^{\mu\nu}}
\def\mN{_\mu^\nu}

\def\wh{wormhole}
\def\whs{wormholes}

\def\ssph{static, spherically symmetric}
\def\asflat{asymptotically flat}
\def\GR{general relativity}

\def\oR{\overline{R}}
\def\og{\overline{g}}

\def\M{{\mathbb M}}
\def\R{{\mathbb R}}

\def\cK{{\cal K}}

\def\ME {\mbox{$\M_{\rm E}$}}
\def\MJ {\mbox{$\M_{\rm J}$}}

\def\eff{_{\rm eff}}

\bls{0.985}
\begin{document}
\twocolumn[

\Title
   {Notes on wormhole existence in scalar-tensor and $F(R)$ gravity}

\Aunames{
    K.A. Bronnikov\auth {a,b,1}, M.V. Skvortsova\auth {b},
    and A.A. Starobinsky\auth {c}
        }

\Addresses{
\addr a {Center of Gravitation and Fundamental Metrology,
       VNIIMS, 46 Ozyornaya St., Moscow 119361, Russia}
\addr b {Institute of Gravitation and Cosmology,
       PFUR, 6 Miklukho-Maklaya St., Moscow 117198, Russia}
\addr c {Landau Institute for Theoretical Physics of RAS,
        2 Kosygina St., Moscow 119334, Russia}
        }

\Abstract
  {Some recent papers have claimed the existence of \ssph\ \wh\ solutions to
  gravitational field equations in the absence of ghost (or phantom)
  degrees of freedom. We show that in some such cases the solutions in
  question are actually not of \wh\ nature while in cases where a \wh\ is
  obtained, the effective gravitational constant $G\eff$ is negative in some
  region of space, i.e., the graviton becomes a ghost. In particular, it is
  confirmed that there are no vacuum \wh\ solutions of the Brans-Dicke theory
  with zero potential and the coupling constant $\omega >-3/2$, except for
  the case $\omega =0$; in the latter case, $G\eff <0$ in the region beyond
  the throat. The same is true for \wh\ solutions of $F(R)$ gravity: special
  \wh\ solutions are only possible if $F(R)$ contains an extremum at which
  $G\eff$ changes its sign.
  %%\PACS {04.50.+h, 04.70.-s, 95.36.+x}
  }

%%\bigskip

] %%%%%%%%%%%%%%%%%%%%%%%%%%%%%%%%%%%%%
\email 1 {kb20@yandex.ru}

\section{Introduction}

  It is well known that to build a static traversable \wh\ in \GR\ it is
  necessary to invoke a matter source of gravity that violates the Null
  Energy Condition (NEC), at least in the neighborhood of the \wh\ throat
  \cite{HV97}. With dynamic \whs\ the situation is more complex: first, the
  notion of a \wh\ throat is then less evident and even admits different
  definitions \cite{HV98,hayw}; second, a dynamic \wh\ may exist not eternally
  but only in a certain time interval, and in this case the requirements to
  its matter source may be weakened \cite{NED-wh}. In what follows, we will
  restrict ourselves to static \wh\ space-times.

  The nonexistence theorem for static \whs\ in the presence of any matter
  respecting the NEC was recently generalized \cite{we07} to the class of
  theories of gravity whose space-times are related to that of \GR\ by a
  conformal mapping. This class includes theories without ghost fields even
  though many of them admit NEC violation. The generalization occurs under
  certain conditions. Thus, for a scalar-tensor theory (STT) of gravity,
  formulated in a space-time \MJ\ (the Jordan frame) with the metric $g\mn$
  using the Lagrangian
\bearr
    L_{\rm STT}= \Half \Big[ f(\Phi) R + h(\Phi)g^{\mu\nu}
            \Phi_{,\mu} \Phi_{,\nu} - 2U(\Phi) \Big]
\nnn \inch
            + L_m     \label{L}
\ear
  ($R$ is the Ricci scalar, $L_m$ is the matter Lagrangian, $f$, $h$ and $U$
  are arbitrary functions), the above theorem holds if the non-minimal
  coupling function $f(\Phi)$ is everywhere positive (in other words, the
  graviton is not a ghost) and also
\beq                                \label{l}
        l(\Phi) := fh + \frac{3}{2}\biggl(\frac{df}{d\Phi}\biggr)^2 >0
\eeq
  (the $\Phi$ field itself is not a ghost).\footnote
       {Our conventions are: the metric signature $(+{}-{}-{}-)$; the
        curvature tensor $R^{\sigma}{}_{\mu\rho\nu} =
        \d_\nu\Gamma^{\sigma}_{\mu\rho}-\ldots,\ R\mn =
        R^{\sigma}{}_{\mu\sigma\nu}$, so that the Ricci scalar $R > 0$ for
        de Sitter space-time and the matter-dominated cosmological epoch; the
        system of units $8\pi G = c = 1$.}  
  The latter condition becomes
  evident if one performs the standard conformal mapping to the Einstein
  frame (the space-time \ME\ with the metric $\og\mn$) such that
\beq
    \og\mn = |f(\Phi)| g\mn,                             \label{conf}
\eeq
  after which the Lagrangian (\ref{L}) transforms to
\bearr
        2L_E = (\sign f) \bigl[\oR
               + \eps \og\MN \phi_{\mu}\phi_{,\nu}\bigr] - 2|f|^{-2}U
\nnn  \inch
                    + 2 |f|^{-2}L_m,            \label{LE}
\ear
  where $\eps = \sign l(\Phi)$, bars mark quantities obtained from or with
  $\og\mn$, indices are raised and lowered with $\og\MN$ and $\og\mn$, and
  the $\Phi$ and $\phi$ fields are related by
\beq                                                            \label{phi'}
        \frac{d\phi}{d\Phi} = \frac{\sqrt{|l(\Phi)|}}{f(\Phi)}.
\eeq

  The Brans-Dicke (BD) STT is the special case of (\ref{L}) corresponding to
\bearr                                                            \label{BD}
       f(\Phi) = \Phi, \cm h(\Phi) = \omega/\Phi,
\nnn
       l(\Phi) = \omega + 3/2, \cm \omega = \const.
\ear

  Let us recall that the NEC reads $T\mn k^{\mu}k^{\nu}\geq 0$, where $T\mn$
  is the stress-energy tensor (SET) of matter and $k^\mu$ is an arbitrary
  null vector with respect to the metric $g\mn$. The mapping (\ref{conf})
  transforms the SET according to ${\overline T}\mn = |f|^{-1} T\mn$ while
  $k^\mu$ remains a null vector in the metric $\og\mn$. Thus if the NEC holds
  in \MJ, it also holds in \ME, and vice versa, and it means, in particular,
  that \whs\ are absent in \ME.

  Assuming that matter in \ME\ does respect the NEC, the above theorem is
  proved in \cite{we07} (following the idea expressed earlier in \cite{vac5})
  using the fact that if both $f(\Phi)$ and $l(\Phi)$ are smooth and positive
  everywhere, including limiting points, the mapping (\ref{conf}) transfers a
  flat spatial infinity in one frame to a flat spatial infinity in the other.
  Therefore, if we suppose that there is an \asflat\ \wh\ in \MJ, its each
  flat infinity has a counterpart in \ME, the whole manifold \ME\ is smooth,
  and we obtain a \wh\ there, contrary to what was assumed. We conclude that
  static and \asflat\ \whs\ are absent in the Jordan frame as well.

  A special case of this situation is connected with matter concentrated on a
  thin shell. Accordingly, as was explicitly shown in \cite{we09}, in any STT
  with a non-ghost scalar field, in {\it any\/} thin-shell \whs\ built from
  two identical regions of \ssph\ space-times, the shell has negative surface
  energy density, thus violating both null and weak energy conditions.

  In \cite{we07}, a possible \wh\ behavior of \ssph\ configurations was also
  considered in STT under weakened requirements, allowing $f(\Phi)$ to reach
  zero or even become negative. It has turned out that if $f$ only reaches
  zero, twice \asflat\ \wh\ solutions in the Jordan frame can exist but only in
  exceptional cases: (i) the corresponding Einstein-frame solution must
  comprise an extreme black hole, whose double horizon is then mapped to
  the second spatial infinity in \MJ; it is not possible with vacuum
  solutions but can happen with nonzero electric or magnetic fields;\footnote
     {Such an example has been found in \cite{BCh07}. In a more general
      context, inclusion of an electric field as a source of gravity in STT
      enlarges the number of classes of solutions but, just as is the case
      with vacuum solutions, \whs\ can exist either with $\eps = -1$ or in
      special cases with $G\eff$ somewhere becoming infinite or negative
      \cite{cold-e}.}
  (ii) additional fine tuning is necessary to avoid a solid angle deficit or
  excess at this second infinity, and (iii) the theory itself should be very
  special.

  Rather a wide (although still special) class of \wh\ solutions exists in
  theories where a transition to $f < 0$ is allowed, so that the manifold
  \ME\ is mapped according to (\ref{conf}) to only a part of the manifold
  \MJ\ (the conformal continuation phenomenon \cite{br73,vac4}). However,
  previous studies have shown that such solutions are generically unstable
  under spherically symmetric perturbations \cite{BG01}, the instability
  appearing due to a negative pole of the effective potential at the
  transition surface to $f < 0$. The existence of this pole still does not
  guarantee an instability, and a further study of non-linear dynamical
  evolution is necessary; but even if such \whs\ might exist, their ``remote
  mouths'' would be located in anti-gravitational regions with $f < 0$. So,
  they cannot connect different parts of our Universe but could only be
  bridges to other universes (if any) with very unusual physics. In
  addition, it is known that a generic space-like anisotropic curvature
  singularity arises dynamically if $f\to 0$ \cite{S81}, and it is unclear
  how to avoid it.

  Some recent publications contain results on \ssph\ \wh\ existence
  which seem to contradict these conclusions. In particular:
\begin{enumerate}
\item
    It is claimed that there are vacuum \wh\ solutions in the BD theory
    corresponding to the coupling constant $\omega$ in the non-ghost
    interval $(-3/2, -4/3)$ \cite{BMIK09}.
\item
    A similar claim is made for the BD theory with $\omega$ in a larger
    non-ghost interval including the value $\omega =0$ \cite{LO10}.
  %%% and even $\omega >0$ \cite{BD++}.
\item
    It is asserted that vacuum BD \whs\ exist for $\omega < -2$ but nothing
    is said about the range $-2\leq \omega < -3/2$ \cite{BMIK09}.
\item
    Wormhole solutions are found in some $F(R)$ theories of gravity
    \cite{LO09} which are equivalent to the BD theory with $\omega=0$
    and a nonzero potential.
\end{enumerate}

  The purpose of this paper is to clarify the situation in all these cases.
  To this end, we begin the next section with designating the conditions
  under which a \ssph\ metric is said to describe a \wh. Then we explicitly
  write the vacuum solution of the general massless STT and specify its
  properties for the BD theory, thus covering items 1--3 above. Section 3
  is devoted to the properties of \wh\ solutions in $F(R)$ theories. Our
  previous conclusions \cite{we07} are confirmed in all these cases.
  Section 4 contains some remarks of methodological nature.

\section {Wormholes in scalar-tensor gravity}
%%%%%%%%%%%%%%%%%%%%%%%%%%%%%%%%%%%%%%%%%%%%%

\subsection{The wormhole notion}

  The general static, spherically symmetric metric can be written as
\beq                                                           \label{ds}
        ds^2 = \e^{2\gamma(u)}dt^2 - \e^{2\alpha(u)}du^2
                                   - r^2(u) d\Omega^2.
\eeq
  where $u$ is an arbitrary radial coordinate and $d\Omega^2 = (d\theta^2 +
  \sin^2\theta d\varphi^2)$ is the linear element on a unit sphere.
  The metric (\ref{ds}) is \asflat\ as $u$ tends to some value $u_0$ (finite
  or infinite) if
\bearr                                                          \label{asflat}
    r\to \infty, \cm \gamma \approx \gamma_0 + \gamma_1/r + o(1/r),
\nnn
    \cm \e^{-2\alpha} r'^2 \to 1,
\ear
  where the prime denotes $d/du$, $\gamma_0,\ \gamma_1 = \const$, and the
  last condition in (\ref{asflat}) is the requirement of a correct
  circumference to radius ratio for large circles.

  We will say that the metric (\ref{ds}) describes a \wh\ if it regular in
  some range $(u_1,\ u_2$) of the radial coordinate, does not contain
  horizons (that is, $\gamma(u)$ and $\alpha(u)$ are finite) in this range,
  and is \asflat\ both as $u\to u_1$ and as $u\to u_2$. The existence of two
  large $r$ asymptotic regions inevitably means that there is at least one
  regular minimum of the function $r(u)$, called a throat.

  We thus consider only globally regular configurations. One could admit more
  general regular asymptotic behaviors, e.g., anti-de Sitter, but for our
  present discussion it is sufficient to be restricted to asymptotic
  flatness. We thus also discard possible wormholes in which $r(u)$ reaches
  large (as compared with the throat radius) but finite values of $r$ at one
  or both sides, as happens, e.g., when a \wh\ connects two closed worlds.
  Our arguments could be easily modified to include such cases.
  Another case of interest which is not covered by our definition and
  deserves a separate study is that of a cosmological horizon located
  far enough from the throat, as is the case in some known models
  \cite{pha1, BLZ10}.

\subsection{Vacuum solutions of a general STT}

  Let us now consider the theory (\ref{L}) assuming $f \geq 0$, $U \equiv 0$
  and $L_m =0$. Then in the Einstein frame we have a massless, minimally
  coupled scalar field as the only source of gravity. The \ssph\ solutions
  to the field equations are well known in this case: these are the Fisher
  solution of 1948 \cite{Fis}) if $\eps = +1$ and the Bergmann--Leipnik
  solution of 1957 \cite{BerLei}) (sometimes called the ``anti-Fisher''
  solution) in case $\eps=-1$. Both solutions were repeatedly rediscovered
  afterwards. Let us reproduce them in the simplest joint form, following
  \cite{br73}.

  Two combinations of the Einstein equations for the metric (\ref{ds}) and
  $\phi = \phi(u)$ read $R^0_0 =0$ and $R^0_0 + R^2_2=0$. Choosing the
  harmonic radial coordinate $u$, such that
  $\alpha(u) = \gamma(u) + 2 \ln r(u)$, we easily solve these equations.
  Indeed, the first of them reads simply $\gamma'' =0$, while the second one
  is written as $\beta'' + \gamma'' = \e^{2(\beta+\gamma)}$. Solving them,
  we have
\bearr
     \gamma = - m u,                                              \label{s}
\nnn
     \e^{-\beta-\gamma} =  s(k,u) := \vars     {
                            k^{-1}\sinh ku,  \ & k > 0, \\
                                 u,          \ & k = 0, \\
                            k^{-1}\sin ku,   \ & k < 0.     }
\ear
  where $k$ and $m$ are integration constants; two more integration constants
  have been suppressed by choosing the zero point of $u$ and the scale along
  the time axis. As a result, the metric has the form \cite{br73}
\beq                                                           \label{ds1}
     ds^2 = \e^{-2mu} dt^2 - \frac{\e^{2mu}}{s^2(k,u)}
                    \biggr[\frac{du^2}{s^2(k,u)} + d\Omega^2\biggl]
\eeq
  (note that flat spatial infinity here corresponds to $u=0$, and $m$ has the
  meaning of the Schwarzschild mass). Moreover, the scalar field equation
  in this gauge reads $\phi''=0$, hence
\beq                                                           \label{phi}
    \phi = C u, \quad\ C = \const \ \ \mbox{(the scalar charge)}
\eeq
  without loss of generality. Lastly, due to the ${1\choose 1}$ component
  of the Einstein equations (the constraint equation), the integration
  constants are related by
\beq                                                           \label{int}
        2k^2 \sign k = 2m^2 + \eps C^2.
\eeq

  \eqs (\ref{ds1}), (\ref{phi}) describe the Fisher solution in the case
  $\eps=+1$, hence $k > 0$ due to (\ref{int}); in the case $\eps = -1$,
  they give the anti-Fisher solution, in which $k$ can be arbitrary. These
  solutions are simultaneously the Einstein-frame vacuum solutions of {\it
  any\/} STT (\ref{L}) with $U(\Phi) = 0$ and $\sign l(\Phi) = \eps$.
  Considering (\ref{ds1}) as $\og\mn dx^\mu dx^\nu$ and applying the mapping
  (\ref{conf}), we easily obtain the solution in the Jordan frame \MJ.

  In particular, in the BD theory we have according to (\ref{phi'})
\bearr
    f(\Phi) = \Phi = \Phi_0\exp (\phi/\sqrt{|\omega+3/2|}),
\nnn
           \Phi_0 = \const,                               \label{Phi}
\ear
  where $\phi = Cu$ and we can put $\Phi_0 = 1$ without loss of generality.

\subsection{The BD theory, $\omega > -3/2$, $\eps = +1$.}

  For $k > 0$, which is always true if $\eps = +1$ (leaving aside the trivial
  case $k=0$ with flat metric), it is helpful to apply the coordinate
  transformation $\e^{-2ku} = 1 - 2k/x $, after which the Jordan-frame
  solution in the BD theory takes the form
\bearr                                                          \label{BD+}
    ds^2_J = P^{-\xi}
         \Big[ P^a dt^2 - P^{-a} dx^2 - P^{1-a} x^2 d\Omega^2 \Big],
\nnn
    \Phi = P^\xi,  \cm   P(x) \equiv 1 - 2k/x,
\ear
  where we have redefined the integration constants:
\beq
       \xi = -C/\big(2k\sqrt{\omega + 3/2}\big), \quad\ a = m/k < 1,
\eeq
  and the relation (\ref{int}) passes on to
\beq                                                            \label{int+}
       (2\omega + 3) \xi^2 = 1 - a^2.
\eeq
  The index $J$ is used to stress that it is the Jordan-frame metric.
  It is the so-called Brans class I solution \cite{B61}, written with the
  explicitly separated conformal factor $1/\Phi$ in the metric; in the square
  brackets in (\ref{BD+}) we have the Fisher metric.\footnote
       {The form (\ref{BD+}) of the solution coincides with the one used
        in \cite{Agn95} if we re-denote $a-\xi=A$, $a+\xi = -B$. The
        substitution $x = y[1 + k/(2y)]^2$ converts it to the isotropic
        coordinates employed, e.g., in \cite{B61, BMIK09}, and the constants
        $C,\lambda$ and $B$ used there are related to ours by
        $C = 2\xi/(a-\xi),\ \lambda = 1/(a-\xi),\ B = k/2$.  \label{fn4}}

  Now, can the solution (\ref{BD+}) describe \whs? To answer this question,
  we notice that the solution is defined and is regular in the range
  $2k < x < \infty$ and is \asflat\ as $x\to\infty$. It is thus sufficient
  to check if the other end, $x = 2k$, can be another flat infinity or a
  regular surface beyond which the solution could be continued.

  The quantity $g_{tt} = P^{a-\xi}$ is finite and non-zero at $x=2k$
  (i.e., $P=0$) only if
  $a = \xi$, i.e., in the special solution in which $g_{tt} \equiv  1$. On
  the other hand, $r^2 = x^2 P^{1-a-\xi}$ is infinite at $x=2k$ if $a + \xi >
  1$ and is finite there if $a + \xi =1$. Thus if $a+\xi >1$, the surface
  area of the sphere $x = \const$ has a minimum (i.e., there is a throat) at
  some intermediate point $x = x_1 > 2k$. (Note that this is a would-be \wh\
  configuration considered in \cite{BMIK09,LO10} in the case $a \neq \xi$.)
  However, in all such cases we have either a naked singularity at $x = 2k$
  or a repulsive non-flat asymptotic unattainable for test bodies.

  Indeed, one can check that, as $x \to 2k$ ($P \to 0$), the Kretschmann
  invariant $\cK = R_{\mu\mu\rho\sigma} R^{\mu\mu\rho\sigma}$ of the metric
  (\ref{BD+}) behaves as $P^{2(a+\xi-2)}$. Hence in case $a +\xi < 2$
  we have a naked singularity at $x=2k$. This happens irrespective of $a$
  being equal to $\xi$ or not, even though $g_{tt} = \const$ when $a=\xi$
  (in which case $\omega < 0$). Thus all such would-be \wh\ configurations
  have naked singularities.\footnote
       {A similar situation occurs in Einstein gravity when one attempts
        to construct a would-be \wh\ solution supported by the phantom
        Chaplygin gas \cite{GKMPS08} or the phantom generalized Chaplygin gas
        \cite{GKMPS09}: a curvature singularity arises at a finite
        value of the spherical radius $r$.}

  For $a + \xi \geq 2$ we have a finite (for $a+\xi =2$) or zero limit of
  $\cK$ at $x=2k$. The range of $\xi$ required for that is
\bearr
  \frac{2-\sqrt{-6\omega -8}}{2\omega
  +4}\le \xi \le \frac{1}{\sqrt{2\omega + 3}},
\nnn \inch\cm
                      -3/2< \omega < -11/8,
\nnnv
  \frac{2-\sqrt{-6\omega -8}}{2\omega +4}\le \xi \le
  \frac{2+\sqrt{-6\omega -8}}
  {2\omega +4},
\nnn \inch\cm
                       -11/8 \le \omega \le -4/3.
\ear
  Thus the maximum possible value of $\omega$ in this case is $\omega=-4/3$,
  achieved at $a=1/2,\ \xi = 3/2$. However, since $a < 1$ and therefore
  $\xi > a$, we inevitably have $g_{tt} = P^{a-\xi} \to \infty$, i.e., this
  asymptotic is repulsive and inaccessible to timelike geodesics. Moreover,
  the limit $x\to 2k$ is characterized by an infinite proper distance along
  the radial direction ($\int P^{-(\xi+a)/2}dx $ diverges). However, the
  iterated integral of the Riemann tensor components in an orthonormal frame
  diverges as $x\to 2k$, just as it happens in the case of usual ``strong''
  singularities, which also indicates the absence of a globally regular
  behavior there (see also the discussion in \cite{Q00} in this connection).
  Summarizing, this configuration cannot be called a \wh\
  according to our definition. Though, one certainly cannot exclude that
  its part containing a throat might be used for obtaining a \wh\ by,
  say, a cut-and-paste procedure with some \asflat\ space-time.

  Let us note that throats appear at all values of $\omega > -3/2$, even
  very large ones. Indeed, suppose, e.g., $\xi = 2(1-a)$, so that $a +\xi=
  2-a >1$. Then by (\ref{int+}) we have
\[
     2\omega + 3 = \frac{1+a}{4(1-a)},
\]
  which can be made arbitrarily large by taking $a$ close enough to unity.
  More generically, as follows from Eq.\,(\ref{int+}), $a+\xi >1$ if
  $0 < \xi < (\omega+2)^{-1}$. But, as we have seen, in all such cases \whs\
  are not obtained.

  Of interest is the special case $a = \xi =1/2 \ \then \ \omega=0$, the only
  case where the sphere $x=2k$ is regular. The metric (\ref{BD+}) then
  acquires the so-called spatial Schwarzschild form
\bear                                                      \label{spa-S}
     ds_J^2 \eql dt^2 - dx^2/(1 - 2k/x) - x^2 d\Omega^2
\nn
            \eql dt^2 - 4(2k + y^2)dy^2 - (2k + y^2)^2 d\Omega^2,
\ear
  where $y^2 = x-2k$. It is a \wh\ metric, defined in the range $y \in \R$.
  The initial range $x > 2k$ (in which the manifold \ME\ is defined)
  corresponds to either $y > 0$ or $y < 0$. The Jordan-frame manifold \MJ\
  consists of two regions $y > 0$ and $y < 0$, each of them mapped into \ME\
  according to (\ref{conf}), plus the regular sphere $y=0$, the throat.

  Thus the only \wh\ solution in this family exists for $\omega = 0$ due to
  the conformal continuation phenomenon \cite{br73, vac4}. The transition
  sphere corresponds to $\Phi = 0$; in the whole range of $y$ we have $\Phi =
  y/\sqrt{y^2 + 2k}$, so that at $y < 0$, beyond the throat, the quantity
  $f(\Phi) = \Phi$ in the Lagrangian (\ref{L}) is negative, which means that
  the effective gravitational constant $G\eff$ is negative there.

  The \wh\ nature of this solution for $\omega = 0$ was pointed out in 1996
  in \cite{br96}, where its multidimensional generalization was also
  indicated.

\subsection{General STT, $\eps = -1$}

  If $\eps = -1$ (so $l(\Phi) < 0$ and $\omega<-3/2$) there are three
  branches in the vacuum solution (\ref{ds1})--(\ref{int}) according to the
  sign of $k$, see (\ref{s}). They correspond to Brans classes II-IV
  \cite{B61}. Of interest for us is the case $k < 0$ (see \cite{br73, cold08}
  for more complete descriptions). Then, as is easily verified, the metric
  (\ref{ds1}) in \ME\ [where now $s(k, u) = k^{-1} \sin ku$] has two flat
  spatial infinities at $u =0$ and $u = \pi/|k|$ (choosing this half-wave of
  the sine function without loss of generality). It is the anti-Fisher \wh,
  repeatedly described beginning with the papers \cite{br73} and
  \cite{h_ell}. And it is also evident that the \wh\ nature of this solution
  is preserved in {\it any\/} STT (\ref{L}) in which the function $f(\Phi)$
  is smooth and positive in the range $0 \leq u \leq \pi/|k|$. This fact was
  already pointed out in 1973 \cite{br73}. The BD theory is just a special
  case of such a theory, see \eq (\ref{Phi}) where, as before, $\phi = Cu$.
  Let us note that the value $\omega = -2$ is not distinguished in any way:
  \wh\ solutions do appear for any $\omega < -3/2$, in the whole ghost range
  of the BD theory.

  For full clarity, let us write this BD solution in a form in which its \wh\
  nature is more obvious. Substituting $|k|u = \cot^{-1}(x/|k|)$, we obtain
\bearr                                                          \label{BD-}
    ds_J^2 = \e^{-(2m + \sigma) u} dt^2 - \e^{(2m-\sigma) u}
                \bigl[ dx^2
\nnn \inch
                + (k^2 + x^2) d\Omega^2 \bigr],
\nnn
        \sigma := -C/\big(2|k|\sqrt{|\omega + 3/2|}\big),
\nnn
        1 + m^2/k^2 = \sigma^2 |2\omega + 3|,
\ear
  where $x$ ranges over the whole real axis. Since $u \in (0, \pi/|k|)$, the
  exponentials in (\ref{BD-}) do not affect the qualitative nature of the
  metric.

\section {Wormholes in $F(R)$ gravity}
%%%%%%%%%%%%%%%%%%%%%%%%%%%%%%%%%%%%%%%

  In the above examples, we only discussed massless field configurations,
  with $U(\Phi) \equiv 0$. However, the theorem proved in \cite{we07}
  concerns the general case of the theory (\ref{L}), with any potentials
  $U(\Phi)$. It therefore encompasses not only STT but also the metric
  $F(R)$ theories of gravity which are known to be equivalent to the BD
  theory with $\omega=0$ and potentials $U(\Phi)$ whose form depends on the
  choice of the function $F(R)$ (see, e.g., the reviews \cite{SaFa08, deFe10}
  and references therein). Indeed, the field equations
\beq    \nq\,                                                   \label{eq-F}
    F_R R\mN - \Half F \delta\mN
       + (\nabla_\mu \nabla^\nu - \delta\mN \Box)F_R - T\mN =0
\eeq
  ($F_R := dF/dR$) obtained by variation of the Lagrangian
\beq                                            \label{LF}
        L_F = \Half F(R) + L_m
\eeq
  (where $F_{RR} := d^2 F/dR^2$ is not identically zero) with respect to
  the metric are equivalent to those due to the Lagrangian (\ref{L}) with
\bearr
       f(\Phi) = F_R, \cm  h(\Phi) =0,
\nnn \inch
          2U(\Phi) = R F_R - F,
\ear
  which is nothing else but the BD theory with $\omega = 0$, whose form
  (\ref{BD}) is obtained by introducing the parametrization $f(\Phi) = \Phi$.
  It should be stressed that the two equivalent forms of the theory
  correspond to the same (Jordan) conformal frame and employ the same metric
  $g\mn$. Thus the condition $f(\Phi) > 0$ takes the form $dF/dR>0$ in
  $F(R)$ gravity. The function $l(\Phi)$ (\ref{l}) cannot become negative,
  however, it can reach zero at exceptional values of $R$ where $F_{RR} = 0$
  while $f = F_R \neq 0$. In the latter case, the parametrization
  $f(\Phi) =\Phi$ is no longer applicable since $f(\Phi)$ loses
  monotonicity.

  Hence, according to \cite{we07}, the only opportunity of obtaining \wh\
  solutions in the theory (\ref{LF}), both vacuum and supported by matter
  respecting the NEC, is connected with conformal continuations through
  a sphere at which $F(R)$ has an extremum reached at some value $R = R_0$.
  As is the case with STT, such spheres, being regular in the Jordan
  frame, are singular in the Einstein frame. After crossing such a sphere,
  one gets \cite{BCh05} to a region with negative values of $\Phi$ in the
  BD formulation of the same theory, which in turn correspond to negative
  values of the effective gravitational constant $G\eff$. It has been shown
  \cite{vac4} that the \wh\ behavior is generic among such special solutions.
  However, similarly to the case $f(\Phi_0)=0$ in STT, in dynamic solutions
  there appears a generic spacelike anisotropic curvature singularity where
  the Kretschmann invariant diverges as $R\to R_0$ \cite{N73,GS79}.

  The opportunity of $F_{RR}=0$ while $F_R \ne 0$ at some $R=R_0$ is also of
  interest. At such surfaces, a weak singularity can arise at which the
  space-time metric and Riemann tensor components are finite (see
  \cite{ABS09} for a discussion of its structure in the cosmological case),
  but $R$ generically behaves there as $R_0 + \lambda \sqrt{u-u_0},\
  \lambda = \const$ if $u = u_0$ is the surface at which $R=R_0$, where $u$
  is the time coordinate in the cosmological setting or a spatial coordinate
  in a static space-time. One can check that the terms with derivatives of
  $F_R$ in \eq (\ref{eq-F}) remain finite at $u=u_0$ in this case. Still
  differential curvature invariants are infinite at such a surface, and there
  is no analytic extension across it. Solutions with $\lambda=0$ which
  are regular at $R = R_0$ are not excluded but require strong fine tuning of
  initial or boundary conditions. However, in all such cases, according to
  \cite{we07}, globally regular \wh\ solutions cannot appear as long as $F_R$
  is everywhere positive.

  Possible \wh\ geometries in $F(R)$ gravity have been discussed in
  \cite{FdB05, LO09}. In \cite{FdB05}, only conditions near a throat were
  discussed; we can once again recall that our results \cite{we07}, based on
  conformal mappings, do not rule out the existence of throats. In \cite{LO09}
  a few specific examples of \wh\ solutions in $F(R)$ gravity, supported
  by matter with anisotropic pressure and without a fully specified equation
  of state, were found. However, in the first two examples, Eqs.\,(25), (31)
  of that paper, $F_R$ is negative over some range of $R$ for matter
  satisfying the weak energy condition. In the other two examples,
  \eqs (35) and (39) of \cite{LO09}, either $F_R$ becomes zero at the throat,
  so that $G\eff$ is infinite and becomes negative beyond it, or the radial
  pressure and/or density of matter diverges there. It is once again
  confirmed that static \wh-like solutions can formally exist in $F(R)$
  theory but look rather unrealistic (cf. \cite{Coule93}).

\section {Concluding remarks}

  One can ask a natural question: why, in rather a simple subject, quite a
  number of incorrect inferences appear in the literature. In our view, one
  reason is purely terminological: some authors call wormholes anything that
  has a throat. Actually their analysis reduces to proving its existence
  only. Meanwhile, to prove that there is a \wh\ it is sufficient to find
  two (not necessarily flat) regular asymptotic regions with growing
  $r(u)$ (or a similar function if spherical symmetry is absent), and the
  existence of its minimum is then evident even without explicitly pointing
  it out.

  Another reason for wrong or incomplete inferences is in some cases a very
  clumsy parametrization used in BD (and other) solutions. For instance, with
  the parameters $C$ and $\lambda$ mentioned in footnote \ref{fn4} it is
  impossible to consider the case $a = \xi$ in the solution (\ref{BD+})
  which is of particular interest.

  And one more oddity of some \wh\ studies is a persistent usage of the
  curvature coordinate $r$, which is certainly convenient for solving the
  field equations in many (but not all) cases but is two-valued in the throat
  region. It really looks funny when a solution is first written in terms of
  an admissible coordinate $u$ but is then transformed to $r$ with much
  effort in order to seek a throat using the so-called shape function $b(r)$.
  Meanwhile, it is sufficient just to find a minimum of $r(u)$. Using $r$
  as a coordinate, one frequently remains restricted to one half of the
  configuration while another half, even if its metric is the same, can have
  different properties (as, e.g., in the solution (\ref{spa-S}) where $G\eff
  < 0$ beyond the throat). In this way, one also sometimes loses \wh\
  solutions which are asymmetric with respect to the throat. Though, with
  certain care, asymmetric \whs\ can certainly be described in the $r$
  parametrization as well \cite{BKim03}.

  We hope that these remarks can be of some use for researchers and
  especially students working in this field.

\Acknow
  {AAS was partially supported by the RFBR grant 08-02-00923 and by the
  Scientific Programme ``Astronomy'' of the Russian Academy of Sciences.
  KB and MS acknowledge partial support from the RFBR grant 09-02-00677a
  and the NPK MU grant of the Peoples' Friendship University of Russia.}

%%\newpage

\end{document}